\documentclass[12pt]{article}
\usepackage{latexsym,epsfig}
\usepackage{graphicx}
\usepackage{pifont}
\usepackage{verbatim}
\def\nn{\nonumber}
\def\ed{\end{document}}

\def\sk{\smallskip}

\def\bg{\bigskip}
\def\beq{\begin{eqnarray}}
\def\eq{\end{eqnarray}}
\def\beqn{\begin{eqnarray*}}
\def\eqn{\end{eqnarray*}}
\def\nl{\noindent}

\begin{document}
\begin{center}
\vskip 1cm
{\bf \large Extra Neutral Gauge Boson}
\bg

{\bf \large from Two Versions of the 3-3-1 Model}
\bg 

{\bf \large  in Future Linear Colliders}
\vskip .8 cm 

{E. Ramirez Barreto, Y. A. Coutinho}

{Universidade Federal do Rio de Janeiro}
\vskip .5 cm
 
{J. S\'a Borges}

{Universidade do Estado  do Rio de Janeiro}
\vskip .5 cm

{Rio de Janeiro - Brazil}
\end{center}

\begin {abstract}

Our aim is to establish some signatures of the extra gauge boson  ${Z^\prime}$, predicted in two versions of ${SU (3)_C \times SU (3)_L \times  U (1)_X }$ model and to show possible differences between these signatures. First, by considering the process  $e^+ + e^- \longrightarrow \mu^+ +\mu^- $, we obtain some observables and next, by considering additional hadronic final states,  we obtain lower bounds for $M_{Z^\prime}$, within $95\%$ C.L.. We also include our preliminary results concerning $p \bar p$ and $p p$ collisions in order to support our conclusions about the possibility to discriminate the various versions of 3-3-1 model.

From our analysis we conclude that linear colliders can show a clear signature for the existence of $Z^\prime$ predicted in the 3-3-1 model and also that it can discriminate the versions.
\end{abstract}
PACS: 12.60.Cn,13.66.De,13.66.Fg,14.70.Pw

elmer@if.ufrj.br, yara@if.ufrj.br, saborges@uerj.br
\newpage
\section{Introduction}
The Standard Model (SM) of the strong and electroweak interactions is extremely successful. Its predictions from calculations both at tree level and for higher order corrections are consistent with all available experimental data. The theoretical extensions of the SM are motivated  by attempting to understand features that are accommodated in the SM but not totally explained by it, for example the family replication problem. Many alternative models address these questions and predict the existence of new fermions and bosons. We are mainly interested in the existence of a new  neutral gauge boson, $Z^\prime$, whose existence is postulated in  3-3-1 models \cite{PIV,FRA,RHN,TON}, little Higgs model \cite{LIT}, left-right symmetric models \cite{LRM}, the superstring inspired E$_6$ model \cite{E6M}, and models with extra dimensions like Kaluza-Klein excitations of neutral gauge bosons \cite{RIZ}. Moreover, the intrinsic properties of this particle and its couplings to matter is strongly model dependent, and it can be used to choose the preferred alternative to the Standard Model.

On the experimental side, the new collider generation will explore the TeV energy regime, opening the possibility of new findings, to be accommodated in a specific  alternative model. The $p p$ Large Hadron Collider (LHC) accelerates protons that can acquire large amounts of energy that can in its turn generate very massive new particles. The associated disadvantage is related to the composite nature of the colliding  protons. This makes it difficult to precisely measure the properties of the newly created particles.

On the other hand, in the future $e^+ e^-$ colliders (ILC \cite{TES} and CLIC \cite{CLI}) the annihilation generates pure energy, from which new particles can be created. Since the starting conditions of the particle production are very well known, the results are much easier to interpret than the results of proton collisions. The high luminosity linear collider is thus a precision machine with which the properties of new particles  e.g. mass, lifetime, spin and quantum numbers  can be very accurately measured. 
These advantages motivate our choice to explore processes in linear colliders that will complement hadron collider findings in an ideal way. 

We choose to work with the 3-3-1 model because its predictions for new physics occur not far above the scale of electroweak symmetry breaking. Besides, this model presents an interesting issue to the family replication problem by relating the number of colors with the number of families in its anomaly cancellation procedure. 

Our work consists in calculating the total cross sections and some fermion distributions   produced by $e^+ e^- $ collisions at ILC and CLIC energies. For fermion pair production, as studied here, the 3-3-1 model includes a $Z^\prime$ contribution besides the $\gamma$ and $Z$ ones. We take into account the scale dependence of all neutral gauge boson couplings in our calculation, and we use two versions of the model in our analysis. We conclude that it is possible to disentangle the models and to estimate bounds on $M_{Z^\prime}$ from the polarized total cross section and the angular distributions of the final fermion.

It is important to extend this analysis to hadron collisions for two reasons: first, because high energy results from Tevatron are yet available, and second, because LHC will start running this year, and the first results are coming soon. Our detailed analysis of muon pair production at Tevatron ($p \bar p$) and LHC ($ pp$) energies and luminosities is still in progress. Due to the complexity of the structure of hadrons, we choose to study the final pseudo-rapidity distribution of the muon by summing up quark and anti-quark (up, down, strange, charm) contributions to obtain bounds on $M_{Z^\prime}$. We have included, in the present work,  our preliminary results to emphasize the differences between the models.

The paper is organized as follows. In Sect. 2, we summarize the relevant aspects of two versions of the 3-3-1 model. In Sect. 3 we present our results for some observables comparing the SM and 3-3-1 models, and we also obtain bounds on the mass of $Z^\prime$ from $e^+ + e^- \to f + \bar{f}$, where $f$ can be a muon, charm and bottom. In this section we also show our preliminary results for bounds on $M_{Z^\prime}$ from $p \bar p$ and $ pp$ collisions. Finally, in Sect. 4 we present our conclusions.  

\section{Models}

The 3-3-1 models are gauge theories with a larger symmetry group than the SM. They are based on the semi-simple gauge group $SU(3)_C \otimes SU(3)_L \otimes U (1)_X$ and, as a consequence, they contain new fermions, scalars and gauge bosons, which have not yet been observed experimentally but are expected to occur at energies near the breaking scale of the SM.

This model offers an explanation of the family replication problem by its anomaly cancellation procedure, requiring that the number of fermion families be a multiple of the quark color number. Knowing that the QCD asymptotic freedom condition is valid only if the number of families of quarks is less than five, one concludes that there are three generations. 

In the 3-3-1 model the electric charge operator is defined by 
\beq 
Q = T_3 + \beta T_8 + X I
\label {beta} \eq
\nl where $T_3$ and $ T_8$ are two of the eight generators satisfying the $SU(3)$ algebra,
\beq \left[ T_i\, , T_j\, \right] = i f_{i,j,k} T_k \quad i,j,k =1 .. 8,\eq
\nl  $I$ is the unit matrix and $X$ denotes the $U(1)$ charge.

The electric charge operator determines how the fields are arranged in each representation, and it depends on the $\beta$ parameter.  
Among the possible choices, $\beta = -\sqrt 3$ corresponds to the minimal version of the model, largely explored in phenomenological applications. The choice $\beta = - 1/\sqrt 3$, which avoids exotic charged fields, leads to a model with right-handed neutrinos. In the following subsections we present the main characteristics of these models.

\subsection{Model I}

For $\beta = -\sqrt 3$ \cite{PIV,FRA} the lepton content of each generation ($a = 1, 2,  3$) is: 
\begin{eqnarray}
\psi_{a L} = \left( \nu_{a}, \ \ell_a, \  \ell^{c}_a \right)_{L}^T\ \sim\left({\bf 1}, {\bf 3}, 0 \right), 
 \end{eqnarray} 

\nl where $\ell^c_a$ is the charge conjugate of the $\ell_a$ (i. e. the $e$, $\mu$ and $\tau$) field. Here the values in the parentheses denote quantum numbers relative to $SU(3)_C$, $SU(3)_L$ and $U(1)_X$ transformations.  

The two quark families ($m=1, 2$) and the third one are accommodated in the $SU(3)_L$ anti-triplet and triplet representation respectively in order to cancel anomalies:
\begin{eqnarray}
Q_{m L} = \left( d_m, \  u_m, \ j_m
\right)_{L}^T \ \sim \left({\bf 3}, {\bf 3^*}, -1/3 \right),  
\quad Q_{3 L} = \left( u_3, \ d_3, \  J
\right)_{L}^T \ \sim \left({\bf 3}, {\bf 3}, 2/3 \right) 
\end{eqnarray}
\begin{eqnarray}
u_{a R}\ \sim \left({\bf 3}, {\bf 1}, 2/3 \right)&,& \  d_{a R} \ \sim \left({\bf 3}, {\bf 1}, -1/3 \right),\nonumber \\
 J_{R}\ \sim \left({\bf 3}, {\bf 1}, 5/3 \right)&,& \  j_{m R} \ \sim \left({\bf 3}, {\bf 1}, -4/3 \right),
\end{eqnarray} 
\nl $j_1$, $j_2$ and $J$ are exotic quarks with respectively, charges $-4/3$, $-4/3$ and $5/3$, in positron charge units.

This version has five additional gauge bosons beyond the SM ones. These are a neutral {$Z'$} and four heavy charged bileptons, ${Y^{\pm\pm},V^\pm} $ with lepton number {$L = \mp 2$}.  

The minimum Higgs structure  necessary for symmetry breaking and giving the quarks and leptons acceptable masses is composed of three triplets  ($\chi$, $\rho$, $\eta$) and one anti-sextet ($S$). 
The neutral scalars of each triplet develop non-zero vacuum expectation values ($v_\chi$, $v_\rho$, $v_\eta$, and $v_S$) and the breaking of the 3-3-1 group to the SM is produced by the following hierarchical pattern
 $${SU_L(3)\otimes U_X(1)}\stackrel{<v_\chi>}{\longrightarrow}{SU_L(2)\otimes
U_Y(1)}\stackrel{<v_\rho,v_\eta, v_S>}{\longrightarrow}{ U_{e.m}(1).}$$
The consistency of the model with SM phenomenology is imposed by fixing a large scale for  $v_\chi$, responsible for the exotic particle masses  ($v_\chi \gg v_\rho, v_\eta, v_S$), with $v_\rho^2 + v_\eta^2 + v_S^2= v_W^2= \left( 246 \right)^2$ GeV$^2$.  

The VEVs induce $Z$-$Z^{\prime}$ mixing. As a consequence the physical states $Z_1$ and $Z_2$ are related to the $Z$ and $Z^{\prime}$ states by the mixing angle,
\beq
\tan^2 \theta = \frac{M^2_Z - M^2_{Z_1}} {M^2_{Z_2} - M^2_Z},
\eq
\nl where $Z_1$ corresponds to the neutral gauge boson of the SM and $Z_2$ to the extra neutral one. 
For small mixing, $\theta \ll 1$, $Z_2$ corresponds to $Z^\prime$.

In this version the relation between the $Z^\prime$ and $Y$ masses \cite{DION,NGL} is
\beq
&&{M_{Y}\over M_{{Z^{\prime}}}} \simeq {M_{V}\over M_{{Z^{\prime}}}} \simeq {\sqrt{3-12\sin^2\theta_W}\over {2\cos\theta_W}}. 
\eq
\nl Using $\sin^2\theta_W =0.23$ \cite{PDG}, we see that this ratio is $\simeq 0.3 $, so that $Z^{\prime}$ can decay to a bilepton pair.
   
The $Z$ and $Z^{\prime}$ interactions involving fermions follow from
\beq
{\cal L}=  -  \frac{g}{2\cos\theta_W}\sum_f\bigl\lbrace \bar
\Psi_f\, \gamma^\mu\ (g_{v f} - g_{a f}\gamma^5)\ \Psi_f\, Z_\mu+ \bar
\Psi_f\, \gamma^\mu\ (g^\prime_{v f} - g^\prime_{a f}\gamma^5)\ \Psi_f\, { Z_\mu^\prime}
\bigr\rbrace, \eq
\nl where $f$ can be leptons or quarks and the vector and the axial couplings,  
$ g_{v f},\ g_{a f}, 
g^\prime_{v f}$ and $ g^\prime_{a f}$, are given in Table I. 

One of the main features of the model comes from the relation between the $SU_L(3)$ and $U_X(1)$ couplings, expressed by
\begin{equation}
\frac {g^{\prime\, 2}}{g^2} =\frac{\sin^2 \theta_W}{1\, -\, 4 \sin^2 \theta_W},
\end{equation}
which fixes $\sin^2 \theta_W < 1/4$, which is a peculiar characteristic of this model.

\subsection{Model II}

The other model considered in the present work corresponds to the choice $\beta = -1/\sqrt 3$ in Eq. (\ref {beta}). It has an anti-neutrino in each $SU(3)_L$ lepton triplet representation \cite{RHN}
\beq
\Psi_{a L} = \left( 
               \nu_a,\  e_a,\  \nu^C_a
 \right)_L^T \sim ({\bf 1}, {\bf 3}, -1/3),\ \  e_{a R}\sim ({\bf 1},
{\bf 1}, -	1), 
\eq
 where $a=1,2,3$ is the generation index.

The two quark generations ($m=1,2$) belong to the anti-triplet and the third ($m=3$) to the triplet representations 
 \beq Q_{m L} = \left(d_{m},\  u_{m},\   d^{\prime}_{i} \right)_L^T \sim ({\bf 3}, {\bf 3^*}, 0), 
\ \ 
 Q_{3L} = \left( u_{3}, \ d_{3}, \ u^\prime_3 
                 \right)_L^T \sim ({\bf 3}, {\bf 3}, 1/3),
\eq
\beq
 u_{a R} \sim ({\bf 3}, {\bf 1}, 2/3),  \ \  d_{a R}\sim ({\bf 3}, {\bf 1}, -1/3),\nn \\
 u_{3 R}^\prime  \sim ({\bf 3}, {\bf 1}, 2/3), \ \ d^\prime_{m R}\sim ({\bf 3}, {\bf 1}, -1/3).
\eq

For symmetry breaking one needs only three triplet scalars ($\chi$, \ $\eta$, \ $\rho$).
Also in this version, for consistency with the low energy phenomenology, the vacuum expectation values obey the condition $v_\chi \gg v_\rho, v_\eta $. The consequences of considering more than three non-zero VEVs are leptonic number violation, Majorana neutrinos mass generation and the existence of a Goldstone boson, called a majoron. 

One of the main features of the model comes from the relation between the $SU_L(3)$  
coupling, $g$, and the $U_X(1)$ coupling, $g^{\prime}$ expressed by
\beq
\frac {g^{\prime \, 2}}{g^2} =\frac{2\sin^2\theta_W}{1 - 4/3 \sin^2\theta_W}.
\eq

The relation between the masses of $Z^\prime$ and the bileptons ($V^\pm$ and $X^0$) \cite{DION,NGL} in the right-handed neutrino model is
\beq
&&{M_{V}\over M_{{Z^{\prime}}}}\simeq {M_{X}\over M_{{Z^{\prime}}}}\simeq{{\sqrt{3-4\sin^2\theta_W}}\over {2\cos\theta_W}}, 
\eq
\nl which will be used in the present work. From this relation we obtain $M_X \simeq 0.82 M_{Z^\prime}$ and then $Z^\prime$ is forbidden to decay into a bilepton pair ($V^+ V^-$ or $X^*  X$).
\sk

The $Z$ and $Z^{\prime}$ interactions involving fermions follow from
\beq
{\cal L} =  -  \frac{g}{2\cos\theta_W}\sum_f\bigl\lbrace \bar
\Psi_f\, \gamma^\mu\ (g_{v f} - g_{a f}\gamma^5)\ \Psi_f\, Z_\mu+ \bar
\Psi_f\, \gamma^\mu\ (g^\prime_{v f} - g^\prime_{a f}\gamma^5)\ \Psi_f\, { Z_\mu^\prime}
\bigr\rbrace, \eq
\nl where $f$ can be leptons and quarks, and the vector and the axial couplings, 
$ g_{v f},\ g_{a f}, 
g^\prime_{v f}$ and $ g^\prime_{a f}$, are given in Table I. 

\section{Results}

\par
Our first comparison between the previously discussed versions is based on the calculation of the total width of the $Z^\prime$. We take two values for the exotic quark masses and we keep the constraints for the $Z^\prime$ and bilepton masses given by Eqs. (7) and (14) for Models I and II, respectively. 

As mentioned before, one of the characteristics of Model I is related to the existence of an upper bound for the Weinberg mixing angle at some scale $\mu$, $\sin^2 \theta_W (\mu) \le 1/4$, which is essential to keep the validity of the perturbation calculation in this model (Eq. (9)).  
The fact that the experimental value of $\sin^2 \theta_W (M_Z)$ is close to $1/4$ ($0.233$) leads to an upper bound associated with the spontaneous symmetry breaking originating in $v_\chi$, which implies directly a restriction for $M_{Z^\prime}$. 

Using just the degrees of freedom corresponding to the SM spectrum, the author of \cite{NGL} concluded that $M_{Z^\prime} < 3.1$ TeV. We respect this constraint by adopting $M_{Z^\prime}$ far below $3.1$ TeV, and we use exotic quark masses with $M_Q \le 1$ TeV.

We display the total $Z^{\prime}$ widths for Models I and II in Figure 1, where one can observe that the widths are one order of magnitude apart. 
Figure 2 shows the dominant branching ratios for $Z^{\prime}$ in both models as a function of $M_{Z^\prime}$ for $M_Q = 600$ GeV. This scenario does not change when we consider exotic quark masses equal to $1$ TeV. In Ref. \cite{PER} the authors have respected the above constraint and considered two possibilities for the exotic quark masses: $M_Q = 500 $ GeV  $ < M_{Z^\prime}$ and $M_Q =  M_{Z^\prime}$. Our results for Model I are compatible with their results.

From Table I we observe that in Model I the $Z^\prime$-leptons couplings are reduced by a factor $\left( 1 - 4 \sin^2 \theta_W\right)^{1/2}$ and the same factor enhances its couplings to the quarks \cite{PER, DUC}. This leads to its {\it leptophobic} character. Moreover, $Z^\prime$ is allowed to decay into a pair of bileptons by Eq. (7) with the same strength.
On the contrary, $Z^\prime$ in Model II couples to the leptons and quarks without any reducing or enhancing factor. 
In Model I, the $Z^{\prime}$ decay modes are dominated by hadronic channels (ordinary and exotic quarks) and bileptons, while in Model II, besides hadronic modes, the charged leptons, neutrinos and Higgs bosons contribute as well. The $Z$ couplings to the fermions in both models are identical to the corresponding SM ones.

The different properties of the extra neutral gauge boson in each model are evident in the calculation of the total cross sections for the processes $e^+ + e^- \longrightarrow \mu^+ + \mu^- $, where $Z^{\prime}$ gives an important contribution together with $\gamma$ and $Z$. 
 
The detector acceptance is such that both muons are detected in the range $\vert \cos \theta \vert \le 0.95 $ \cite{OPA}. Since our results are not sensitive to this angular cut, we adopt $\vert \cos \theta \vert \le 0.995$. Figure 3 shows the resulting total cross sections against $\sqrt s$ for the SM and the models considered for two values of the exotic quark masses ($600$ GeV and $1$ TeV) and for $M_{Z^\prime} = 800$ GeV and $2$ TeV. In both panels the sharp peaks on the left come from the $Z$ contribution, and the SM background is represented by a continuous line.  

For both $Z^\prime$ masses, and for exotic quark masses equal to $600$ GeV, the  total cross section of Model I is similar to the SM one, and it is almost flat around the $Z^\prime$ poles. On the other hand, the total cross section of Model II presents a sharp peak. For $\sqrt s \simeq 800 $ GeV and for an annual luminosity ${\cal L} = 100$ fb$^{-1}$, the number of events expected for Model II is $\simeq 10^6$, while for Model I it is $\simeq 5\times 10^4$. For $\sqrt s \simeq 2 $ TeV, the number of events decreases to $\simeq 10^5$ for Model II and to $\simeq 5\times 10^3$ for Model I. For $M_{Z^\prime}= 2$ TeV the total cross section peak becomes higher and narrower for higher exotic quark masses ($1$ TeV). From now on we restrict our study to exotic quark masses equal to $600$ GeV. In summary, these results show that the total cross sections themselves would reveal the signature of new physics in the resonance region, and this observable can distinguish the two versions.  

At this point it is interesting to discuss the scale dependence of our results. From Table I we observe that all neutral gauge boson couplings to the fermions depend on $\sin^2 \theta_W$. It is known that this parameter gets its scale dependence from the renormalization group equations satisfied by the electroweak couplings $g$ and $g^{\prime}$ \cite{ALG}. We observe that Model I couplings are more sensitive to rapid running of the scale. For example, at $\sqrt s \simeq 800 $ GeV ($\sin\theta_W \simeq 0.488$) \cite{ALG}, $\Gamma_{Z^\prime}= 222$ GeV, to be compared with $142$ GeV for $\sin\theta_W(M_Z) \simeq 0.48076$. Despite this large correction, the total cross section for  $e^+ + e^- \longrightarrow \mu^+ + \mu^- $ decreases from $\sigma= 0.252$ pb to $\sigma =0.182$ pb, becoming closer to the SM result $\sigma_{SM}=0.174$ pb. This effect is smaller for higher energy, and it is related to the Breit-Wigner shape of the $Z^\prime$ contribution to $\sigma$. For Model II the change of $\sigma$ when including the scale dependence can reach a maximum of $2.5\%$ at the resonance region. 
This discussion leads us to perform all calculations with scale dependent couplings.

Furthermore, in order to enhance the sensitivity of linear colliders to distinguish the models, we analyze the total cross section for muon pair production when the beams are longitudinally polarized. In Figure 4 we present the total cross section $\sigma_{LR}$ for $\sqrt s= 1$ TeV when the electron is left-handed polarized and the positron is right-handed polarized.  We observe that Model II, in a region around the resonance, is more sensitive to polarization effects than Model I (coincides with SM), allowing one to use this observable to extract bounds on $M_{Z^\prime}$, as will be discussed later.

Another observable in the same process is the forward-backward charge asymmetry $A_{FB}$ which is defined by the angular distribution of the muon with respect to the electron direction. This observable is sensitive to the extra gauge boson having different couplings to the leptons, as shown in Figure 5, which presents $A_{FB}$ {\it versus} $M_{Z^\prime}$ for $\sqrt s =1$ TeV and $3$ TeV. It can be noticed that Model I leads to the same constant asymmetry as the SM one ($A_{FB}^{SM} = 0.48$), but the asymmetry of Model II varies strongly "on" and "off" the $Z^\prime$ peak. This behavior can be explained in: Model II $Z^\prime$ is not {\it leptophobic}, and we conclude that this observable can also distinguish these models.

Left-right asymmetry is an observable that relies on experiments using polarized beams. We take the polarizations of the electron and positron beam as $P_- = - 90\%$ and $P_+ = 60\%$, respectively. Although Model II presents a dependence on the mass of $Z^\prime$, we note in Figure 6 that this observable does not show a meaningful deviation from the SM background. 

Up to now, our analysis is based on the different couplings of the extra neutral gauge boson to the leptons. It is interesting to include $Z^\prime$ couplings to the quarks, for example, one can extract bounds on the $Z^\prime$ mass for each model by including the production of a pair of  charm and bottom quarks. These processes have the advantage that, at tree level, they introduce only a few unknown parameters, since there is no contribution from exotic particles, and also because {\it a priori} they belong to different representations in each model.

Our strategy is to perform a $\chi^2$ analysis of the difference of the angular distribution of the final fermions relative to the initial beam obtained by the SM with those predicted by the two versions of the 3-3-1 model.
The procedure is as follows: supposing the experimental data for fermion pair production to be described by the SM, we define a one-parameter $\chi^2$ estimator
\smallskip
\begin{equation}
{ \chi^2 = \sum_{i=1}^{n_b} {\biggl( {N_i^{SM}- N_i^{331} \over
\Delta N_i^{SM}}\biggr)^2}},
\end{equation}
where $N_i^{SM}$ and $N_i^{331}$  are the number of events collected in the
$ith$ bin for the SM and the 3-3-1 model, respectively, and $\Delta N_i^{SM} =
\sqrt{(\sqrt {N_i^{SM}})^2 + (N_i^{SM}\epsilon)^2}$ is the corresponding
total error, which combines in quadrature the Poisson-distributed statistical
error with the systematic error. 

\par
In our calculation we took $30$ equal-width bins (a large number of events) and the systematic error $\epsilon = 5\%$. We considered the muon, charm and bottom detection efficiency as $95\%$, $60\%$ and $35\%$, respectively.

Using the three referred channels, we estimated the bounds for $M_{Z^\prime}$ with a $95\%$ C.L. for the energy range of $\sqrt s$ from $0.5$ TeV to $2$ TeV, corresponding to the proposed ILC and CLIC experiments. Our results are displayed in Figures 7, 8 and 9 for the $\bar \mu \mu$,  $\bar c c$ and  $\bar b b$ channels, respectively. We observe that the leptonic and quark production channels give quite different bounds for both models and opposite energy behavior. 

Returning to our analysis of the polarized total cross section shown in Figure 4 and keeping the number of bins, the detection efficiency and the systematic error as before, we have obtained 
$M_{Z^{\prime}} > 650$ GeV with $95\%$ C. L. for Model II. On the other hand, it was not possible to establish any bound from the polarized total cross section of Model I. 

As mentioned before, a perturbative treatment in Model I avoiding a Landau-like pole imposes the bound $M_{Z^{\prime}} < 3.1$ TeV when using just the SM particle content. The analysis of the running of $\sin^2 \theta_W(\mu)$ in a recent work \cite {ALE} includes the 3-3-1 degrees of freedom that enlarge this bound to $M_{Z^{\prime}} < 5.2$ TeV, which corresponds to a symmetry breaking scale of $\mu_{331} = 2$ TeV.
 
From the experimental point of view, the absence of a signal in dilepton mass and angular  distributions leads to a bound on the masses of $Z^\prime$ as predicted in many models.   
At the LEP collider ($e^+ e^-$ collisions) the experimental limits for ${Z^\prime}$ mass are $710$, $898$ and $1018$ GeV, obtained respectively by the DELPHI, ALEPH and OPAL collaborations \cite{PDG}. 

There are also bounds coming from hadron colliders. For example, by assuming that $Z^\prime$ couples to ordinary matter with the same strength as the $Z$, the CDF collaboration at Tevatron ($p \bar p$ collisions at $\sqrt s =1.96$ TeV) have obtained  $825$ GeV \cite{PDG}.

As discussed before, we have extended our present analysis to hadron collisions, using again dimuon production. In the case of $p \bar p$ collisions, the $\chi^2$ fit was done for the final muon pseudo-rapidity distribution $\partial \sigma /\partial \eta$ by introducing the following cuts: $m_{\mu \mu} > 200$ GeV, $E_\mu > 5$ GeV and $\vert \eta_\mu\vert\leq 2.5$ GeV, for both muons. Within $95 \%$ C.L. and with a muon detection efficiency of $90\%$ and an integrated luminosity of $340$~pb$^{-1}$ our preliminary results are $M_{Z^\prime} > 620$ GeV and $M_{Z^\prime} > 640$ GeV for Models I and II, respectively, which values are close to the CDF experimental results for the spin one $Z^\prime$ of $E_6$ \cite{CDF}.

It is urgent to present some estimates for $p p$ collisions at LHC,  which will start running  this year, with $\sqrt s =14$ TeV. The discovery limits for an extra gauge boson, considering an integrated luminosity of $100$~fb$^{-1}$, can reach $\simeq 5$ TeV \cite{GOD,DIT}. We have extended our $\chi^2$ fit calculation to this range of energy and luminosity by assuming $80\%$ muon detection efficiency and by applying the same cuts as before to obtain $M_{Z^\prime} > 1700$ GeV and $> 1500$ GeV for Models I and II, respectively, with $95 \%$ C.L.

\par
Let us mention that several theoretical approaches have been used to obtain bounds on the mass of $Z^\prime$ and the mixing angle for the 3-3-1 models. One of these calculates the oblique electroweak correction parameters ($S$, $T$ and $U$) due to the contribution of the exotic particles \cite{LIU,SAS}. Other direct approaches use the experimental results at the $Z$-pole generalized to arbitrary $\beta$ values \cite{OCH,FRE,GUT} or the contributions of $Z^\prime$  to the mass difference of the neutral mesons ($K$, $D$ and $B$) due to the flavor changing neutral current (FCNC) \cite{VAN,DUM,TAE}. The bounds on the mass of $Z^\prime$ were also obtained by considering the energy region where a perturbative treatment is still valid \cite{ALE,PHF}.  

An indirect method to establish limits on the mass of $Z^\prime$ follows from the relation between the masses of $Z^\prime$ and the exotic bosons. It calculates the exotic boson contribution to the muon decay parameters \cite{NGL,BEL}. Besides, in a 3-3-1 model with lepton families in different representations, a bound for the $Z^\prime$ mass from $\mu \rightarrow 3 \ e $ was obtained \cite{SHE}.

\begin{figure} 
\includegraphics[height=.3\textheight]{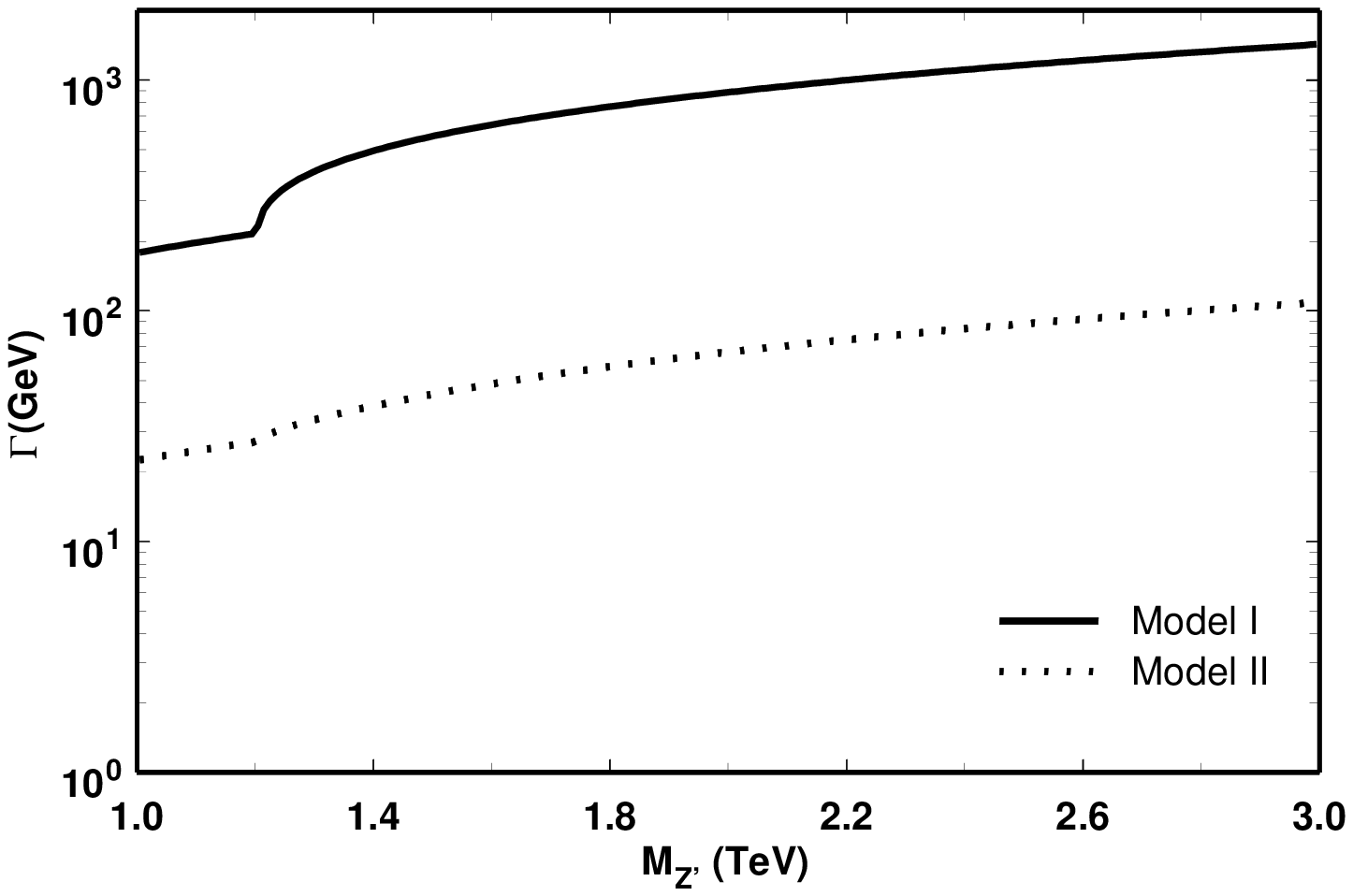}
\hskip -0.5 cm \includegraphics[height=.3\textheight] {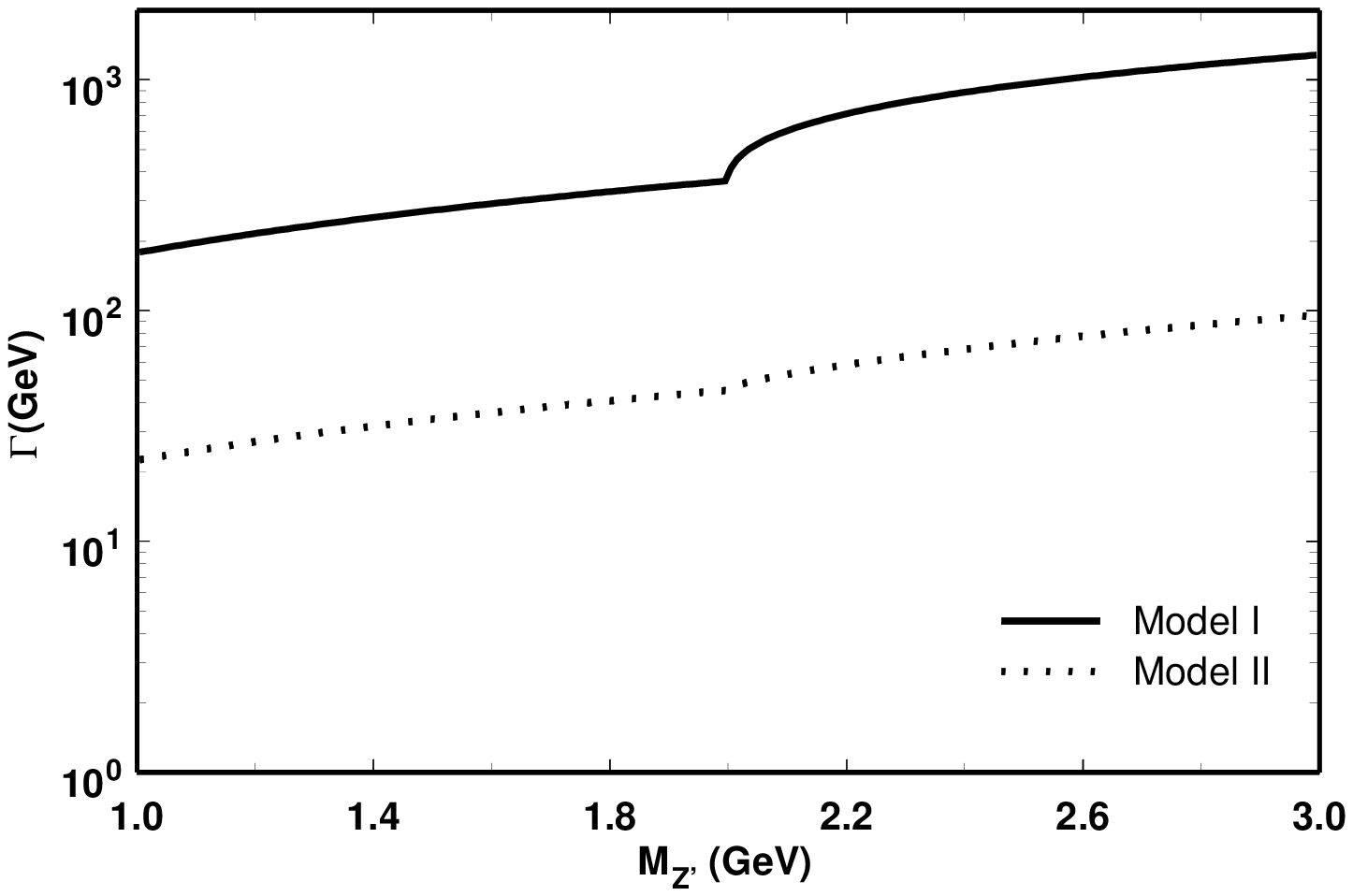}
\caption{The $Z^\prime$ width as a function of $M_{Z^\prime}$ for Models I and II considering exotic quark masses equal to $600$ GeV (left) and $1$ TeV (right).}
\end{figure}

\begin{figure} 
\includegraphics[height=.3\textheight]{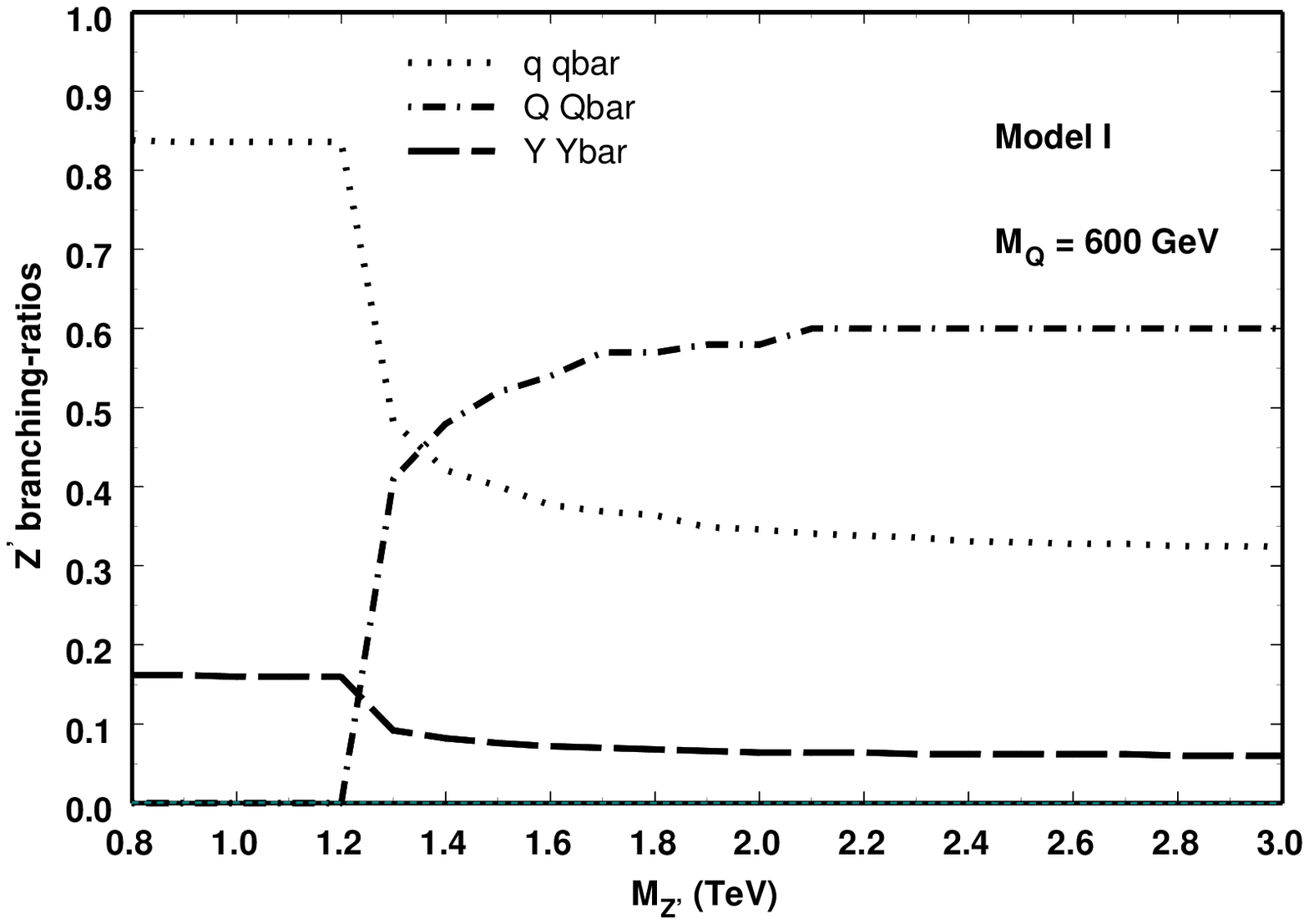}
\hskip -0.5 cm \includegraphics[height=.3\textheight] {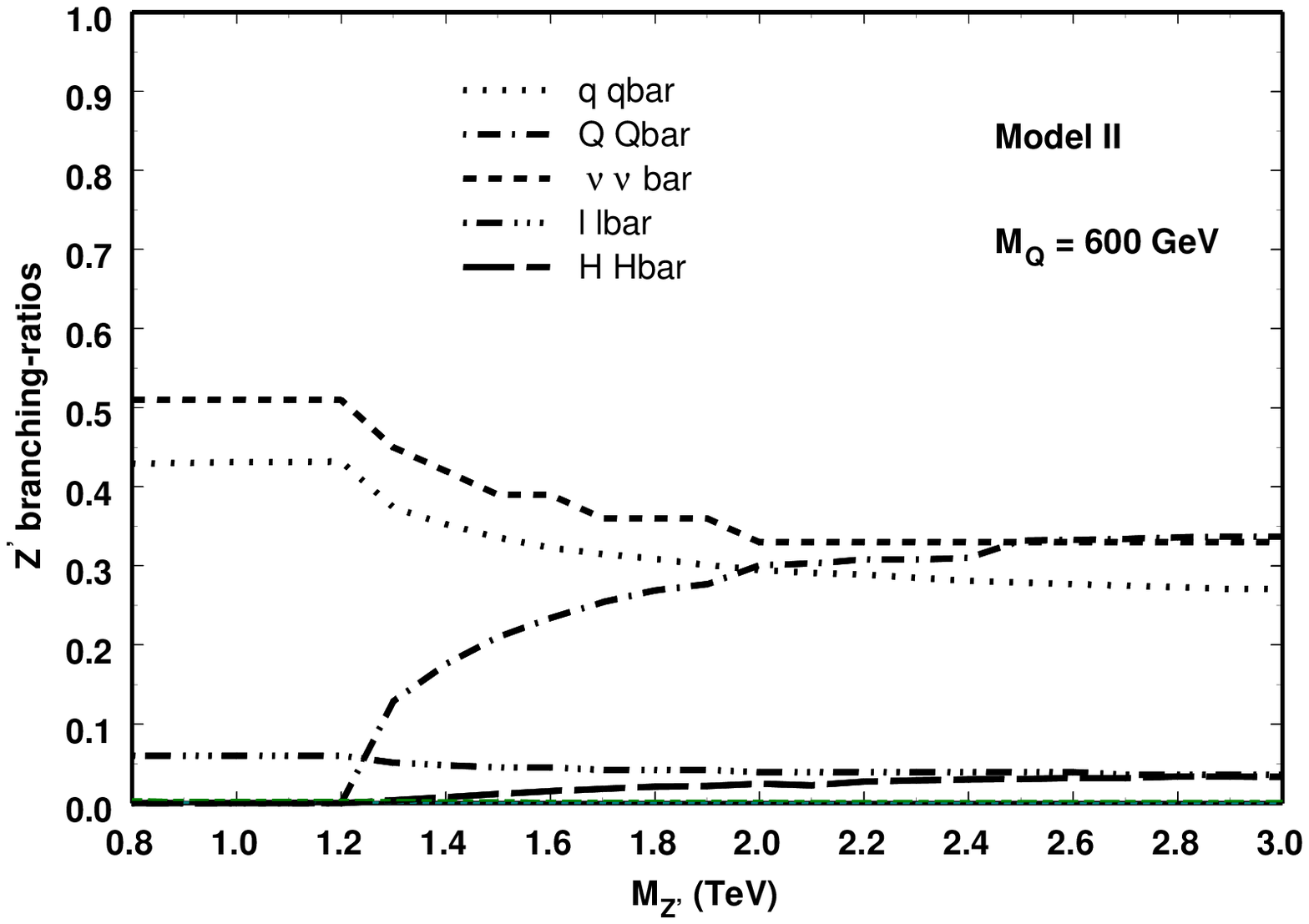}
\caption{The $Z^\prime$ dominant branching ratios for Model I (left) and Model II (right), considering exotic quark masses equal to $600$ GeV.}
\end{figure}

\begin{figure} 
\includegraphics[height=.3\textheight]{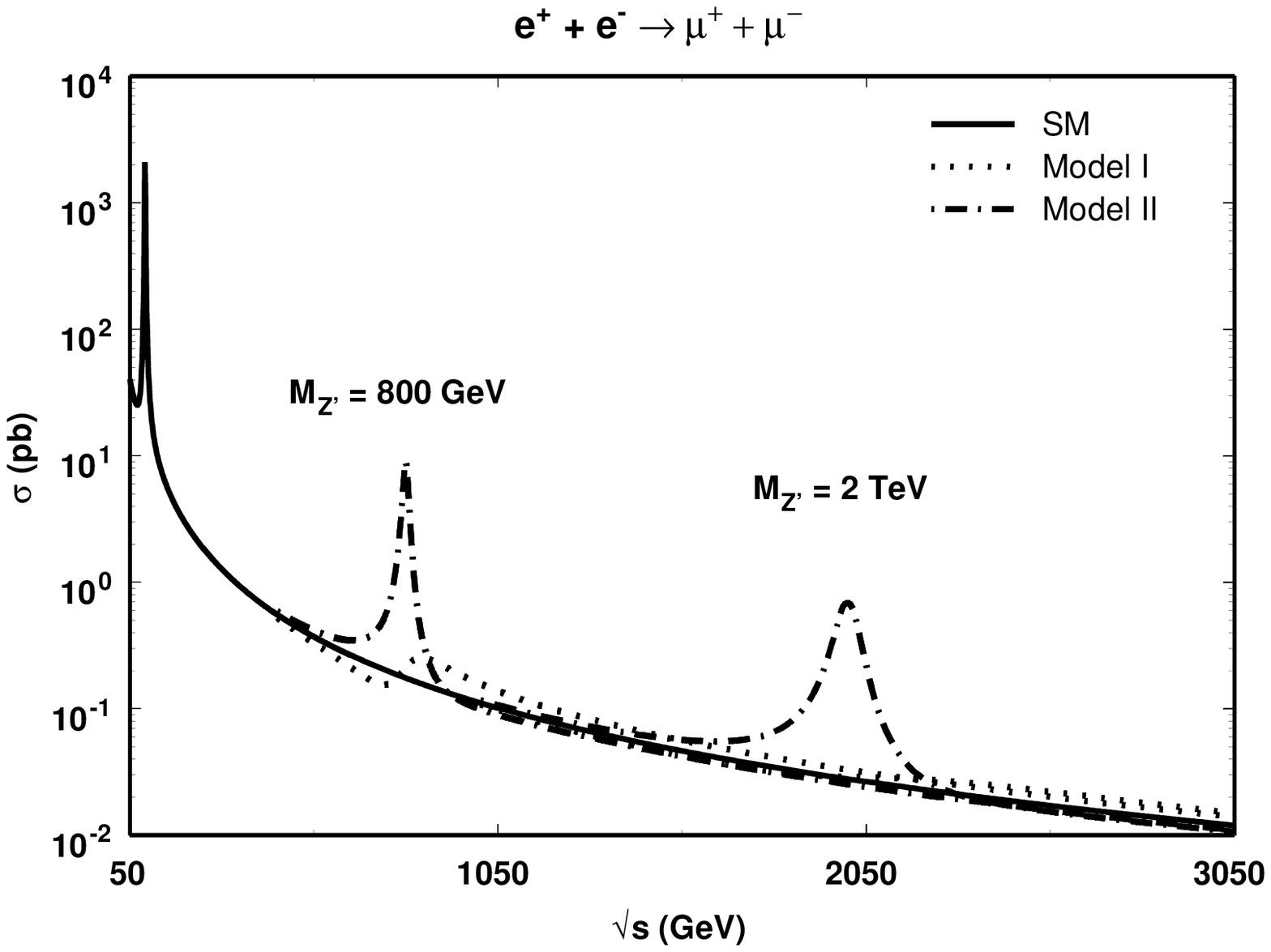}
\hskip -0.5 cm \includegraphics[height=.3\textheight] {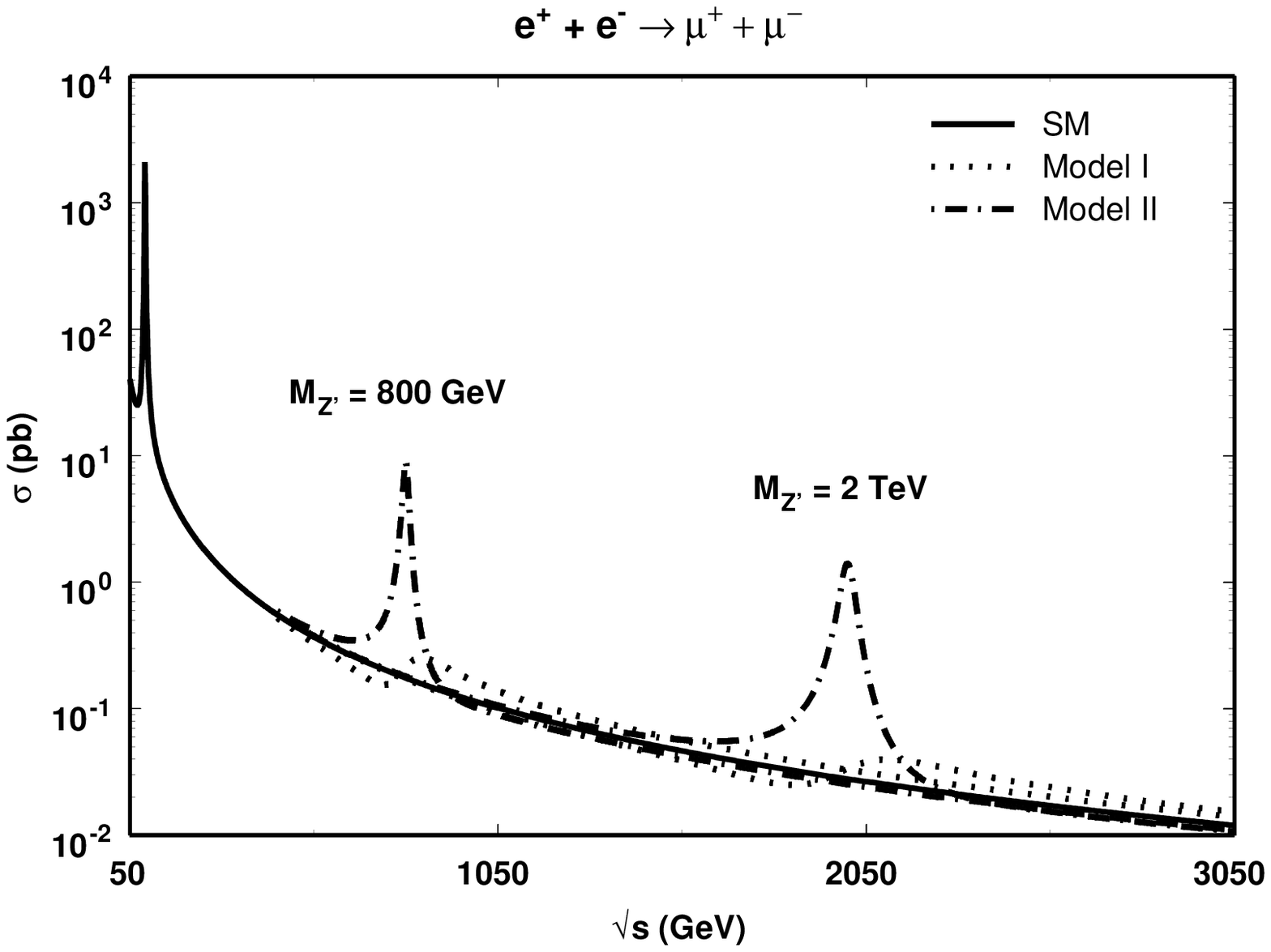}
\caption{The $e^+ + e^- \rightarrow \mu^+ + \mu^-$ total cross sections against $\sqrt s$ for SM, Models I and II for $M_{Z^\prime}= 800$ GeV and $M_{Z^\prime} = 2$ TeV, for exotic quark masses equal to $600$ GeV (left) and $1$ TeV (right).}
\end{figure}

\begin{figure} 
\includegraphics[height=.5\textheight]{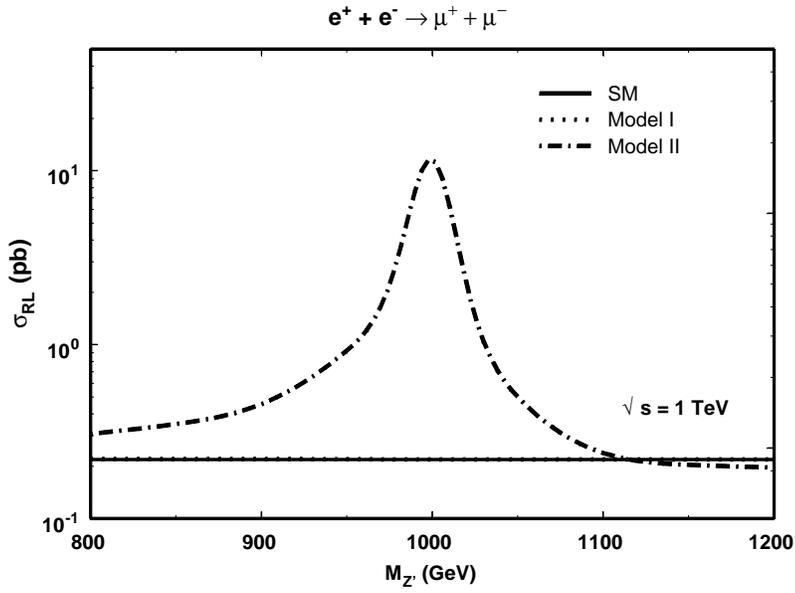}
\caption{The longitudinally-polarized $e^+ + e^- \rightarrow \mu^+ + \mu^-$ total cross section against $M_{Z^\prime}$ for SM and Models I and II for $\sqrt s = 1$ TeV.}
\end{figure}

\begin{figure} 
\includegraphics[height=.3\textheight]{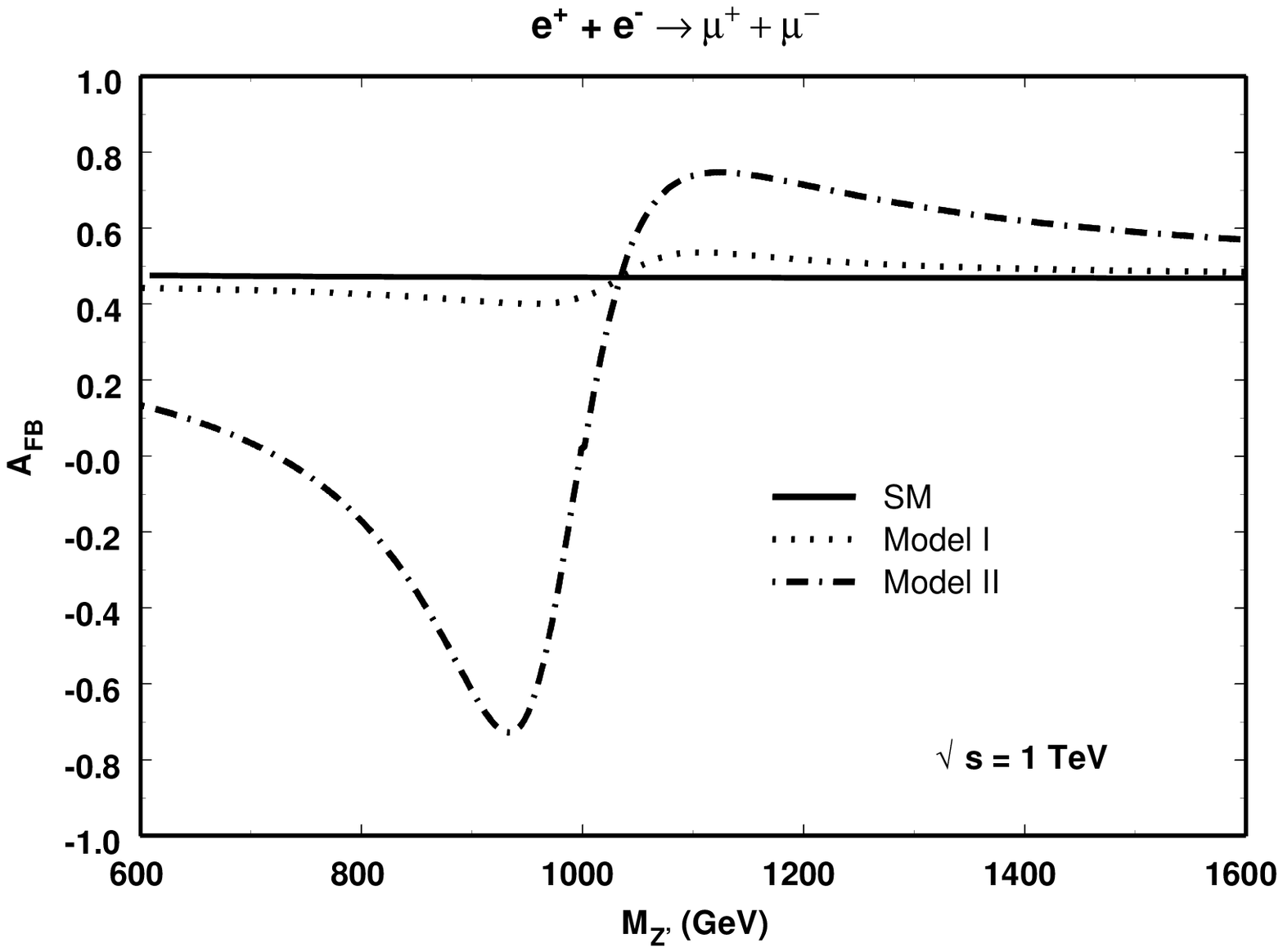}
\hskip -0.5 cm \includegraphics[height=.3\textheight] {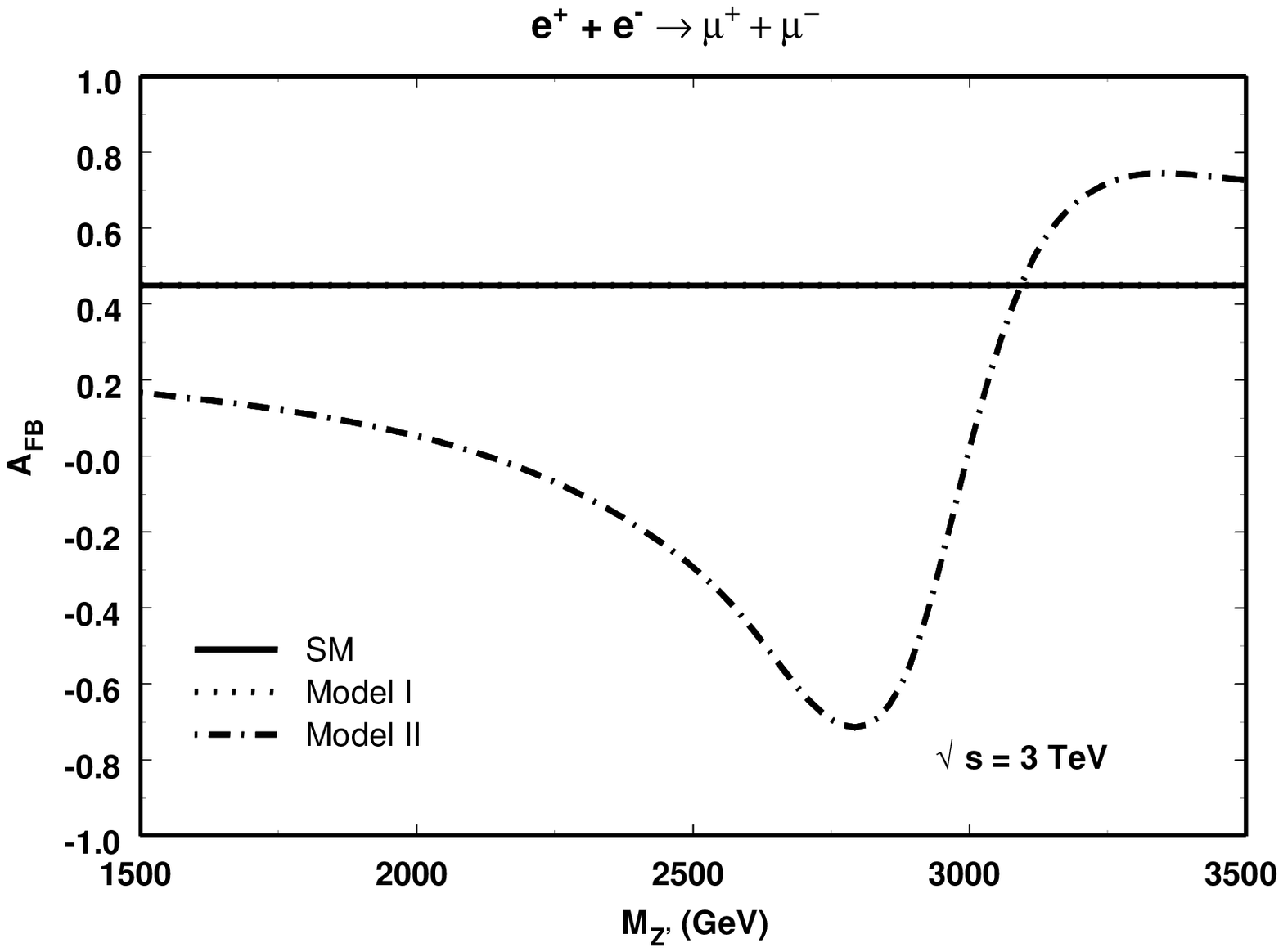}
\caption{The forward-backward asymmetry in the process $e^+ + e^- \rightarrow \mu^+ + \mu^-$ {\it versus} $M_{Z^\prime}$ for SM and Models I and II, for $\sqrt s = 1$ TeV ( left) and  $3$ TeV (right).}
\end{figure}

\begin{figure} 
\includegraphics[height=.3\textheight]{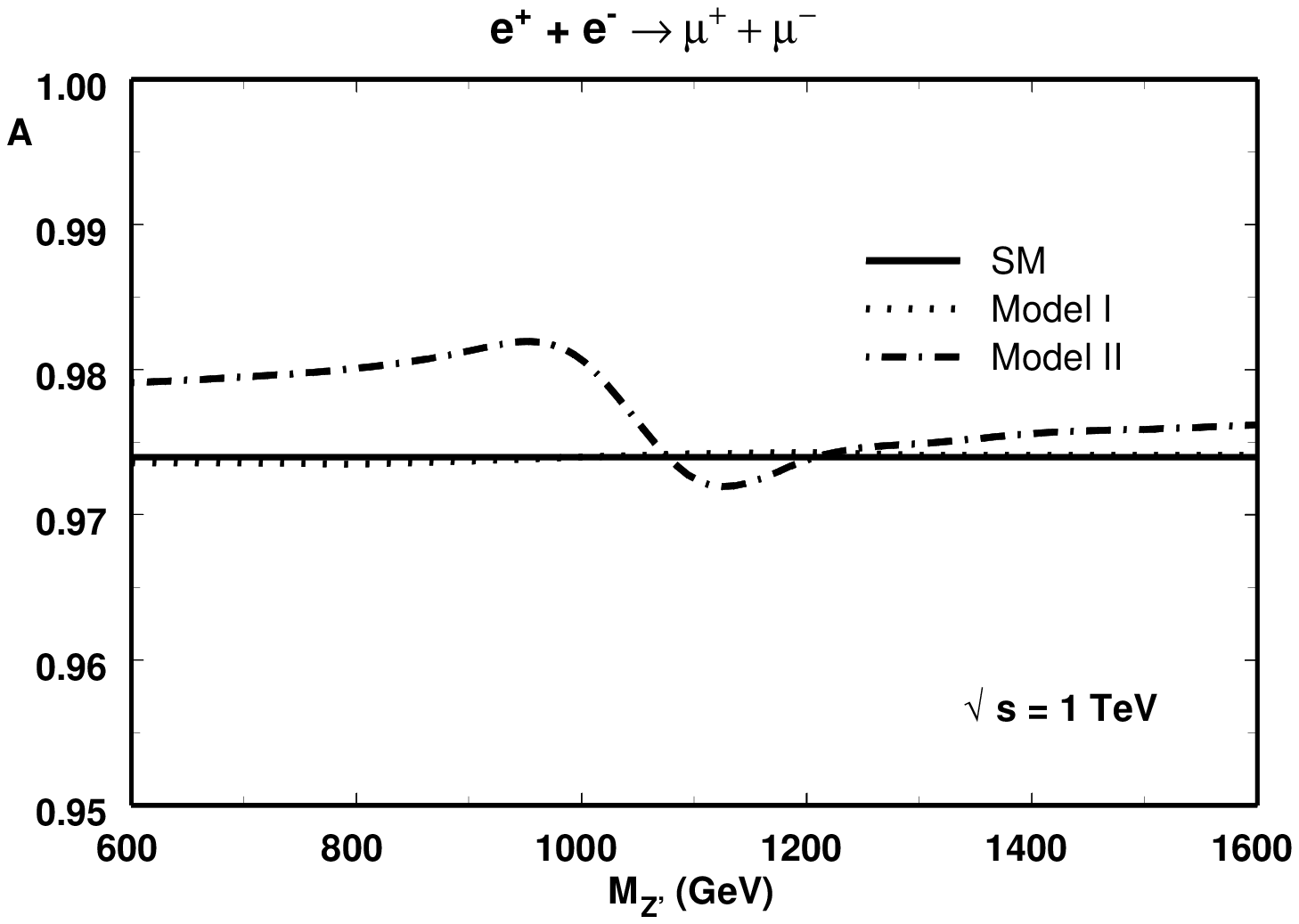}
\hskip -0.5 cm \includegraphics[height=.3\textheight] {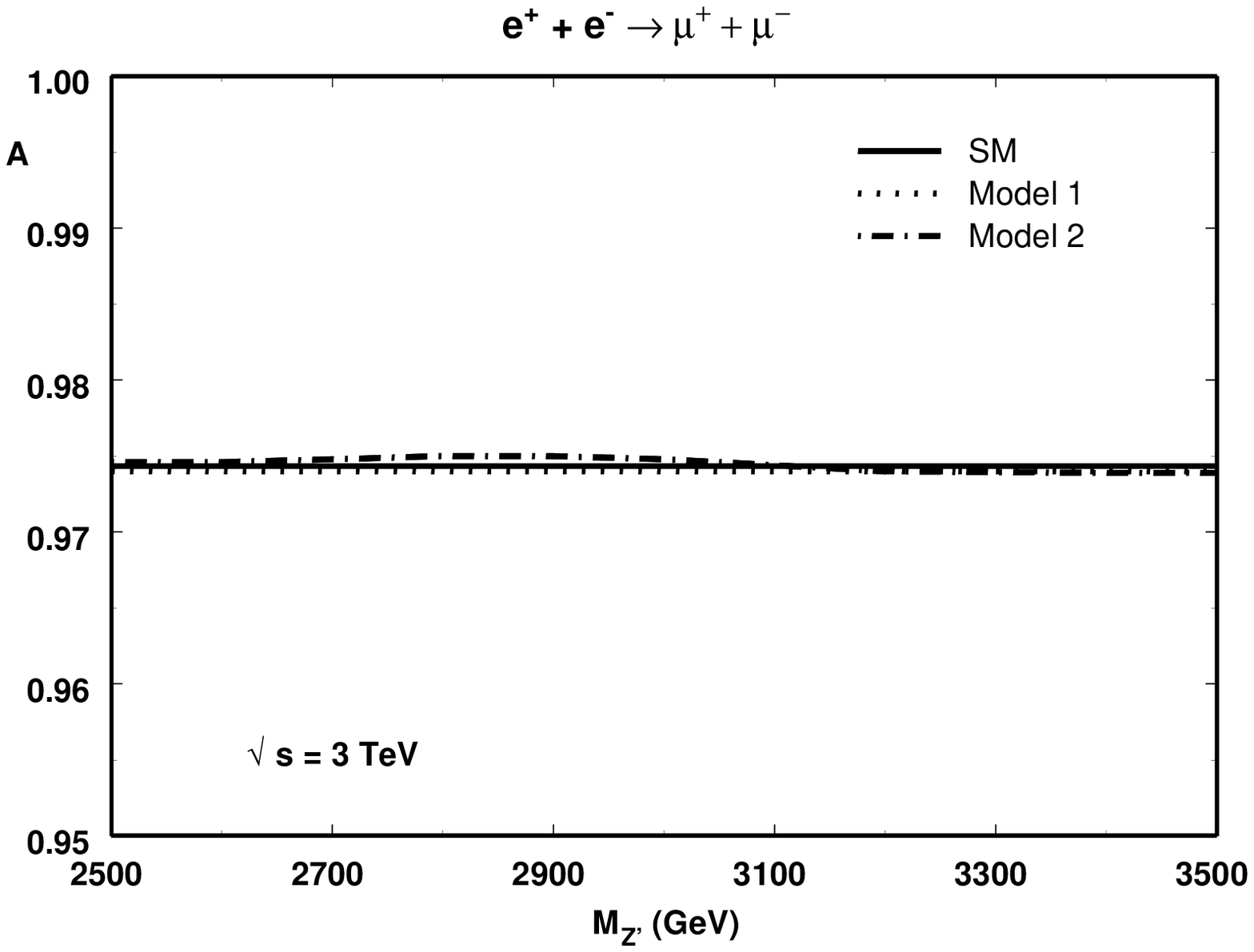}
\caption{The left-right asymmetry in the process $e^+ + e^- \rightarrow \mu^+ + \mu^-$ {\it versus} $M_{Z^\prime}$ for SM and Models I and II, for $\sqrt s = 1$ TeV (left)  and  $3$ TeV (right).}
\end{figure}

\begin{figure}
\includegraphics[height=.5\textheight]{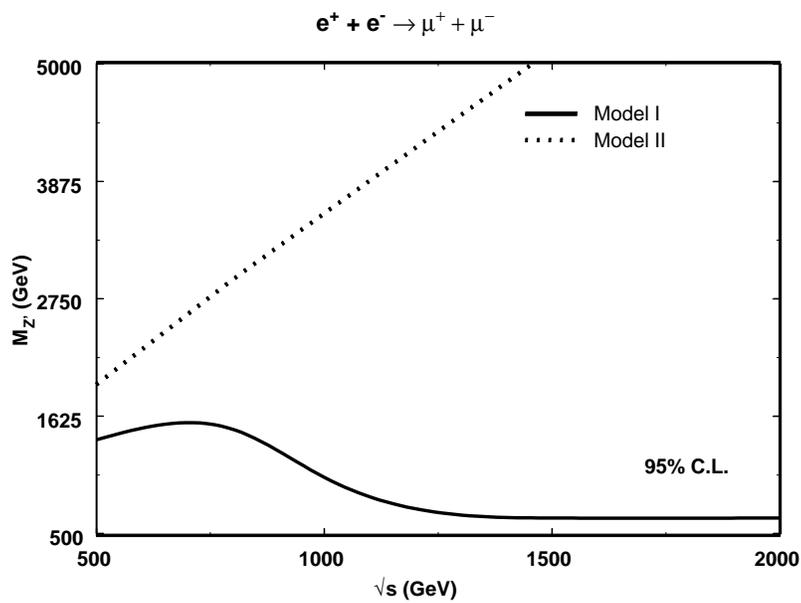}
\caption{The lower bounds at $95 \% $ C.L. extracted from the muon angular distribution  relative to the beam direction {\it versus} $\sqrt s$, for $e^+ + e^- \rightarrow \mu^+ + \mu^-$ for Models I and II.}
\end{figure}

\begin{figure}
\includegraphics[height=.5\textheight]{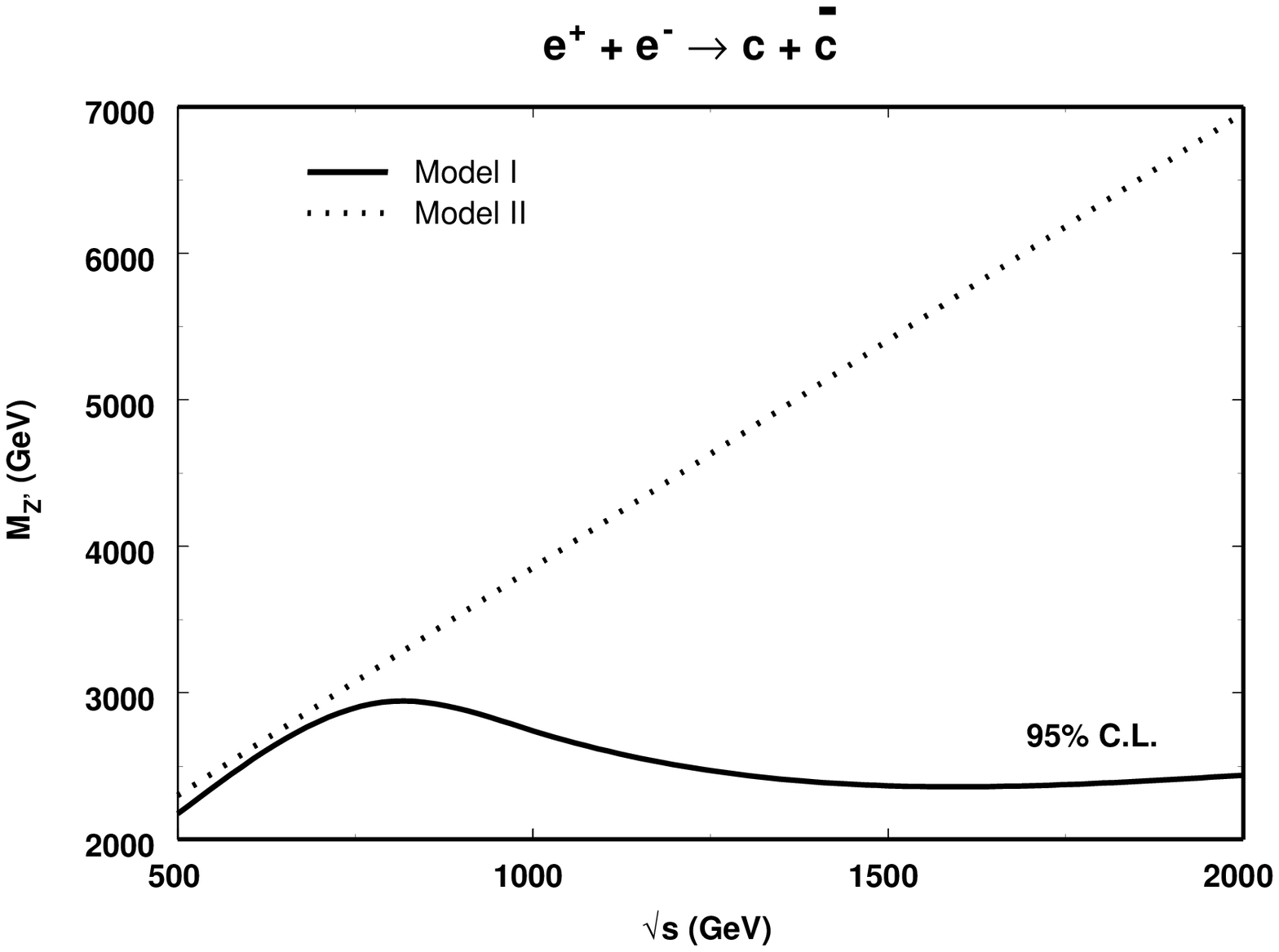}
\caption{The lower bounds at $95 \% $ C.L. extracted from the charm angular distribution  relative to the beam direction {\it versus} $\sqrt s$, for $e^+ + e^- \rightarrow c + \bar c$ for  Models I and II.}
\end{figure}

\begin{figure}
\includegraphics[height=.5\textheight]{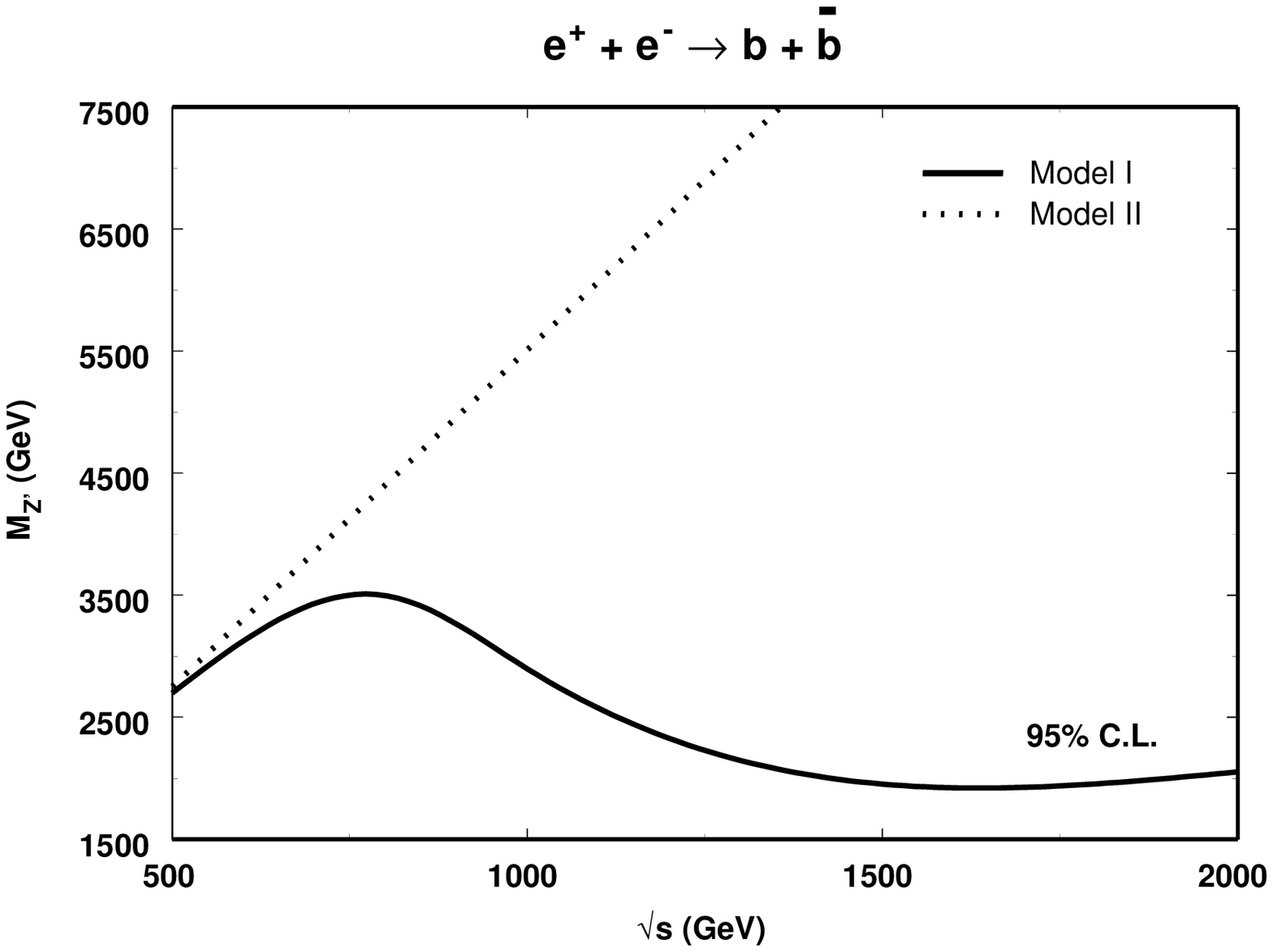}
\caption{The lower bounds at $95 \% $ C.L. extracted from the bottom angular distribution  relative to the beam direction {\it versus} $\sqrt s$, for $e^+ + e^- \rightarrow b +\bar b$ for Models I and II.}
\end{figure}

To be more specific let us compare our results with those obtained from the exhaustive analysis done in Ref. \cite{OCH}. The authors of this reference do a $\chi^2$ fit with $95 \%$ C.L.  between LEP parameters at the $Z$ pole and the predictions from three different quark representations of the 3-3-1 models ($A$, $B$ and $C$). They consider three degrees of freedom: the mass of $Z^\prime$, the mixing angle $\theta$ between the mass states of $Z$ and $Z^\prime$,  and the $\beta$ parameter; they display the allowed region in the $\sin \theta \times \beta $ plane for some $Z^\prime$ masses. In our calculation we used the following representations: C with $\beta=-\sqrt 3$ for Model I and A with  $\beta=-1/\sqrt 3$ for Model II, and we do not consider any mixing between the mass states of $Z$ and $Z^\prime$. The allowed region in the  Ref. \cite{OCH} always includes our zero mixing choice. They also present their allowed region in the $M_{Z^\prime}\times \beta $ plane for  $10 ^{-3}> \sin \theta >10 ^{-4}$. In spite of their small mixing, our results are compatible with their bounds. 
  
\begin{table}[h]\label{sagui}
\begin{footnotesize}
\begin{center}
\begin{tabular}{||c|c|c|c|c||}
\hline \hline
\multicolumn{3}{|c|}{Model I} &
\multicolumn{2}{|c|}{Model II}
\\ 
\hline

&  $g_V$ & $g_A$   & $g^{\prime}_V$ & $g^{\prime}_A$ \\
&    &    &    &       \\ \hline
\hline
&    &     &    &     \\
$Z \bar \nu_l \nu_l$ & $\displaystyle{\frac{1}{2}}$ &  $\displaystyle{\frac{1}{2}}$ & $\displaystyle{\frac{1}{2}}$ &  $\displaystyle{\frac{1}{2}}$   \\
&     &    &    &   \\ \hline
\hline
&     &    &    &   \\
$Z \bar l l$ & $\displaystyle{-\frac{1}{2}+2\sin^2\theta_W}$ &  $\displaystyle{-\frac{1}{2}}$ & $\displaystyle{-\frac{1}{2}+2\sin^2\theta_W}$ & $\displaystyle{-\frac{1}{2}}$\\
&     &    &    &   \\ \hline
\hline
&     &    &    &   \\
$Z \bar c c$ & $\displaystyle{\frac{1}{2}-\frac{4\sin^2\theta_W}{3}}$ &
$\displaystyle{\frac{1}{2}}$ & $\displaystyle{\frac{1}{2}-\frac{4\sin^2\theta_W}{3}}$ &
$\displaystyle{\frac{1}{2}}$ \\
&     &    &    &   \\ \hline
\hline
&     &    &    &   \\
$Z \bar b  b$ & $\displaystyle{-\frac{1}{2}+\frac{2\sin^2\theta_W}{3}}$ &  $\displaystyle{-\frac{1}{2}}$ &  $\displaystyle{-\frac{1}{2}+\frac{2\sin^2\theta_W}{3}}$ &  $\displaystyle{-\frac{1}{2}}$ \\
&     &    &    &   \\ \hline
\hline
&     &    &    &   \\
$Z^{\prime} \bar \nu_l \nu_l$  &   $\displaystyle{-\frac{\sqrt{3}}{6}\sqrt{1-4\sin^2\theta_W}}$  &
$\displaystyle{-\frac{\sqrt{3}}{6}\sqrt{1-4\sin^2\theta_W}}$ &   $\displaystyle{\frac{-1+2\sin^2\theta_W}{2\sqrt{3-4\sin^2\theta_W}}}$ &
$\displaystyle{\frac{-1+2\sin^2\theta_W}{2\sqrt{3-4\sin^2\theta_W}}}$ \\
&     &    &    &     \\  \hline
\hline
&     &    &    &     \\
$Z^{\prime} \bar l l $ &
$\displaystyle{-\frac{\sqrt 3}{2}{\sqrt{1-4\sin^2\theta_W}}}$ &
$\displaystyle{\frac{\sqrt 3}{6}{\sqrt{1-4\sin^2\theta_W}}}$ &
$\displaystyle{\frac{-1+4\sin^2\theta_W}{2\sqrt{3-4\sin^2\theta_W}}}$ &
$\displaystyle{-\frac{1}{2\sqrt{3-4\sin^2\theta_W}}}$ \\
&     &    &    &   \\ \hline
\hline
&     &    &    &   \\
$Z^{\prime} \bar c c$ & $\displaystyle{\frac{\sqrt{3}(1-6\sin^2\theta_W)}{6\sqrt{1-4\sin^2\theta_W}}}$
& $\displaystyle{\frac{-\sqrt{3}(1+2\sin^2\theta_W)}{6\sqrt{1-4\sin^2\theta_W}}}$ &  $\displaystyle{\frac{3-8\sin^2\theta_W}{{6\sqrt{3-4\sin^2\theta_W}}}}$  & $\displaystyle{\frac{-1}{2\sqrt{3-4\sin^2\theta_W}}}$  \\
&    &   &  &   \\ \hline
\hline
&     &    &    &      \\
$Z^{\prime} \bar b b$ &  
$\displaystyle{\frac{\sqrt{3}}{6\sqrt{1-4\sin^2\theta_W}}}$  &  $\displaystyle{\frac{-\sqrt{3}\sqrt{1-4\sin^2\theta_W}}{6}}$  &  $\displaystyle{\frac{-\sqrt{3-4\sin^2\theta_W}}{6}}$   &  $\displaystyle{\frac{1}{2\sqrt{3-4\sin^2\theta_W}}}$  \\
&      &    &    &    \\ \hline
\hline
\end{tabular}
\end{center}
\end{footnotesize}
\caption{The vector and axial couplings of $Z$ and $Z^{\prime}$ with leptons ($\nu_e$ and $e$) and quarks ($c$ and $b$) in Models I and II. $\theta_W$ is the Weinberg angle.}
\end{table}

Let us focus on the production of an on-shell $Z^\prime$ pair with $M_{Z^\prime} = 800$ GeV, having in mind that the extra gauge bosons predicted in a 3-3-1 model have different properties. The resulting total cross sections are displayed in Figure 10, which also includes the dominant SM contribution for two $Z$ production.
From this graph, it is clear that only the CLIC machine can produce a pair of such heavy neutral bosons, but SM production dominates, and it is more than one order of magnitude above the 3-3-1 background for both models. Model II presents a total cross section larger than Model I. For an annual integrated luminosity of ${\cal L}_{int}= 100$ fb$^{-1}$, considering $\sqrt s \simeq 2$ TeV, the number of events with two final $Z^\prime$ is ${\cal O} (10^{2})/yr$ for the 3-3-1 models and ${\cal O} (10^{3})/yr $ for two final $Z$ in SM. For extra gauge bosons heavier than $800$ GeV, the curves continue below the SM one. 

\begin{figure}
\includegraphics[height=.5\textheight]{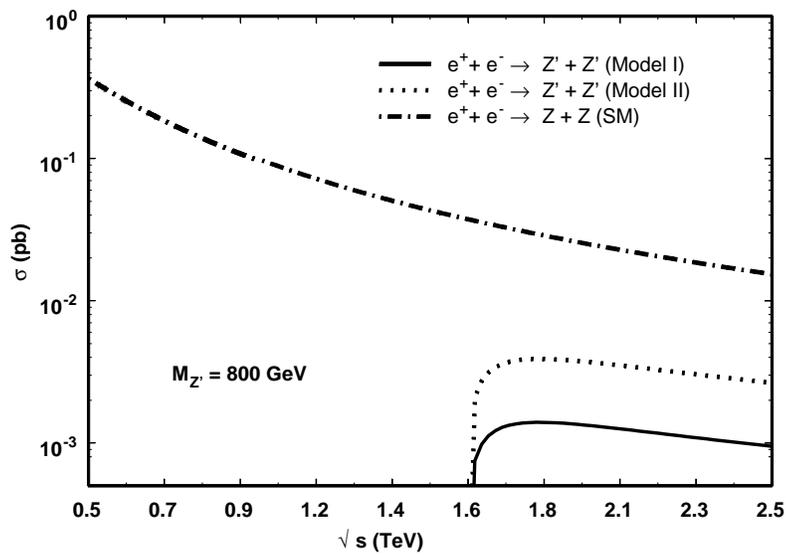}
\caption{Total cross sections for $e^+ + e^-$ collisions producing two neutral gauge boson
against $\sqrt s$ for SM and Models I and II.}
\end{figure}

\section{Conclusions}
Working with two versions of the 3-3-1 model, the minimal version (Model I) and one that allows for right-handed neutrinos (Model II), we explore the energy range accessible at the future linear colliders (ILC and CLIC), $\sqrt s = 0.5$ TeV to $2$ TeV, using scale dependent gauge boson couplings. The first observable that shows differences between these models is the total  width of $Z^\prime$. We conclude that $Z^\prime$ from Model I is broader than $Z^\prime$ from Model II, due to its coupling to the quarks and bileptons. 
This property affects the total cross section for $e^+ + e^- \rightarrow \mu^+ + \mu^- $, which we analyzed for two values of the mass of $Z^\prime$. Model I, similarly to the SM background, leads to a flat distribution around the $Z^\prime$ pole, while Model II presents a sharp peak about two orders of magnitude above the SM, and this peak gets sharper and higher for a heavier exotic quark. This is a clear signature for distinguishing the models. 
We have extended this analysis to longitudinally-polarized beams, and we conclude that this observable enhances the differences between these models.

For the same process we calculate the forward-backward asymmetry that exhibits differences between the models. Model I has the same constant value as the SM ($A^{SM}_{FB} =0.48$), but Model II shows a very strong dependence on $M_{Z^\prime}$ for the "on" and "off" the $Z^\prime$ peak. 
This follows from the very small $Z^\prime$-lepton couplings in Model I in contrast with 
Model II. On the other hand, left-right asymmetry does not present any measurable deviation from the SM for both models.
 
Apart from these signatures we also establish some lower bounds for $Z^\prime$ to be found. We obtained the bounds for $M_{Z^\prime}$ from the final angular distribution of the fermion for $e^+ + e^- \longrightarrow \bar f +  f$, with $f=\mu$, $c$ and $b$. For the channel corresponding to the production of a pair of muons, our results are in agreement with the data from LEP. 
Only for Model II it was possible to extract bounds from  muon pair production in the polarized total cross section. 

We extended the analysis of the contribution of ${Z^\prime}$ in $p \bar p$ at Tevatron and $pp$ collisions at LHC, in order to complement our study and to support our present conclusions in the framework of linear colliders.
Our preliminary estimates for the lower bounds in 3-3-1 models are $620$ GeV (Model I) and 
$ 640$ GeV (Model II) for $p \bar p$, in agreement with Tevatron experiments at $\sqrt s= 1.96$ TeV, and $1500$ GeV (Model I) and $1500$ GeV (Model II) for $p p$ collision at $\sqrt s= 14$ TeV. These latter numbers are included in the expected capability of LHC to reach $Z^\prime$ up to $5$ TeV for ${\cal L} = 100$ fb$^{-1}$.

In this work, our main proposal was to show that, in the 3-3-1 framework, the analysis of $e^+ + e^- $ collisions can reveal a clear signature for the existence of the extra neutral gauge boson. Besides, we have shown that it is possible to disentangle two different versions of the model by the total cross sections and the asymmetries in the angular distribution. Finally, we have established bounds on $M_{Z^\prime}$ from the production of a pair of muons for linear and hadron collisions and of quarks only for linear collisions. 

It could be interesting to compare different versions of 3-3-1 models with identical quark representations or different quark representations for the same model by collision processes such as $e^+ + e^-$ or $p + p$. Such an analysis could reveal the equivalence or non-equivalence of different quark representations.

\vskip 1cm

We acknowledge financial support from CAPES (E.~R.~B.~) and FAPERJ (Y.~A.~C.~). We would like to thank F. M. L. de Almeida Jr. for helpful discussions.

\ed
\begin{thebibliography}{ABC}
\bibitem{PIV} F.~Pisano and V.~Pleitez, Phys. Rev. D {\bf 46}, 410 (1992). 
\bibitem{FRA} P.~H.~Frampton, Phys. Rev. Lett. {\bf 69}, 2889 (1992). 
\bibitem{RHN} J.~C.~Montero, F.~Pisano and V.~Pleitez, Phys. Rev. D {\bf 47}, 2918 (1993); R.~Foot, H.~N.~Long and T.~A.~Tran, Phys. Rev. D {\bf 50}, R34 (1994); Hoang Ngoc Long, Phys. Rev. D {\bf 53},  437 (1996); {\it ibid} {\bf 54}, 4691 (1996); V.~Pleitez, Phys. Rev. D {\bf 53}, 514 (1996).
\bibitem{TON} V.~Pleitez and  M.~D.~Tonasse, Phys. Rev. D {\bf 48}, 2353 (1993). 
\bibitem{LIT} N.~Arkani-Hamed, A.~G.~Cohen and H.~Georgi, Phys. Lett. B {\bf 513}, 232 (2001); N.~Arkani-Hamed, A.~G.~Cohen, E.~Katz and A.~E.~Nelson, JHEP {\bf 0207}, 034 (2002).
\bibitem{LRM} An extensive list of references can be found in  R.~N.~Mohapatra and P.~B.~ Pal, "Massive Neutrinos in Physics and Astrophysics", World Scientific, Singapore, 1998.
\bibitem{E6M} For a review see e.g. J.~L.~Hewett, T.~G.~Rizzo, Phys. Rep. {\bf 183}, 193 (1989). 
\bibitem{RIZ} For a pedagogical review see e.g. T.~G.~Rizzo, e-Conference C040802, L013 (2004).
\bibitem{TES} R.~Brinkmann {\it et al.} (eds.), TESLA Technical Design Report, DESY-2001-011, March 2001; Report from the International Linear Collider Technical Review Committee.
G.~A.~Loew (SLAC). SLAC-PUB-10024, Jul 2003. 
\bibitem{CLI} CLIC Physics Working Group, E.~Accomando {\it et al.}, hep-ph/0412251.
\bibitem{DION} B.~Dion, T.~Gregoire , D.~London, L.~Marleau, H.~Nadeau, Phys. Rev. D{\bf 59}, 075006 (1999).
\bibitem{NGL} Daniel Ng, Phys. Rev. D {\bf 49}, 4805 (1994).
\bibitem{PDG} Particle Data Group, W.-M. Yao {\it et al.}, J. Phys. G {\bf 33}, 1 (2006).
\bibitem{PER} M.~A.~Perez, G.~Tavares-Velasco and J.~J.~Toscano, Phys. Rev. D {\bf 69}, 115004 (2004).
\bibitem{DUC} Le Duc Ninh, Hoang Ngoc Long, Phys. Rev. D {\bf 72}, 075004 (2005). 
\bibitem{OPA} G.~Abbiendi {\it et al.},  OPAL Collaboration,  Eur. Phys. J. C13, 553  (1999).
\bibitem{ALG} A.~G.~Dias, Phys. Rev. D {\bf 71}, 015009 (2005).
\bibitem{ALE} A.~G.~Dias, R.~Martinez, V.~Pleitez, Eur. Phys. J. C {\bf 39}, 101 (2005).
\bibitem{CDF} A.~Abulencia {\it et al.}, Phys. Rev. Lett. {\bf 95}, 252001  (2005).
\bibitem{GOD} S.~Godfrey, Proceedings of APS / DPF / DPB Summer Study on the Future of Particle Physics (Snowmass 2001), Snowmass, Colorado, 344 (2001).
\bibitem{DIT} Michael Dittmar, Anne-Sylvie~Nicollerat, Abdelhak Djouadi, Phys. Lett. B {\bf 583}, 111 (2004).
\bibitem{LIU} James T.~Liu, Daniel Ng, Zeit. Phys. C {\bf 69}, 693 (1994).
\bibitem{SAS} K.~Sasaki, Phys. Lett. B {\bf 308}, 297 (1993). 
\bibitem{OCH} A.~Carcamo, R.~Martinez and F.~Ochoa, Phys. Rev D {\bf 73}, 035007 (2006).
\bibitem{FRE} Fredy Ochoa and R.~Martinez, hep-ph/0508082.
\bibitem{GUT} Diego A.~Gutierrez, William Ponce and Luis A.~Sanchez,  hep-ph/0411077.
\bibitem{VAN} Hoang Ngoc Long and Vo Thanh Van, J. Phys. G {\bf 25}, 2319 (1999). 
\bibitem{DUM} D. G\'omez Dumm, F.~Pisano, V.~Pleitez, Mod. Phys. Lett. A {\bf 9}, 1609 (1994).
\bibitem{TAE} Tae Hoon Lee, Dae Sung Hwang, Int. J. of Mod. Phys. A {\bf 12}, 4411 (1997).
\bibitem{PHF} Paul H.~Frampton, James T.~Liu, B.~Charles Rasco, Daniel Ng, Mod. Phys. Lett. A {\bf 9}, 1975 (1994).
\bibitem{BEL} I.~Beltrami, H.~Burkard, R.~D.~Von Dincklage, W.~Fetscher, H.-J.~Gerber, 
K.~F.~Johnson, E.~Pedroni, M.~Salzmann and F.~Scheck, Phys. Lett. B {\bf 194}, 326 (1987).
\bibitem{SHE} David L.~Anderson and Marc Sher, Phys. Rev  D {\bf 72}, 095014 (2005).
\end{thebibliography}
